\documentstyle[prl,floats,twocolumn,aps,epsf,psfrag,amssymb]{revtex}
\tighten
\begin{document}
\draft
\title{Simulating nonlinear spin models in an ion trap}
\author{G.J. Milburn}
\address{Isaac Newton Institute , University of Cambridge,\\
Clarkson  Road, Cambridge, CB3 0HE, UK.\\
 and Department of Physics, The University of Queensland,QLD 4072 Australia.}
\date{\today}
\maketitle
\begin{abstract}
We show how a conditional displacement of the vibrational mode of trapped ions can be used to simulate
nonlinear collective and interacting spin systems including nonlinear tops and Ising models ( a universal two qubit gate),
independent of the vibrational state of the ion. Thus cooling to the vibrational ground state is unnecessary provided the
heating rate is not too large. 
\end{abstract} 

One of the paths leading to the current interest in quantum computation begins with attempts to answer Feynman's
question\cite{Feynman}: can quantum physics be efficiently simulated on a classical computer? It is generally believed that the
answer is no, although there is no explicit proof of this conjecture. It then follows that a computer operating entirely by
quantum means could be more efficient than a classical computer.  Our belief in this conjecture stems from a number of
algorithms, such as Shor's factorisation algorithm\cite{Shor1994}, which appear to be substantially (even exponentially)
more efficient than the classical algorithms. A number of schemes have now been proposed for a quantum
computer, and some have been implemented in a very limited way.  What kinds of simulations  might these schemes enable? A
number of investigators have attempted to answer this question\cite{Lloyd96,Abrams97,Lidar97,Boghosian98,Zalka98}.  In this
paper we consider this question in the context of the ion trap quantum computer model and show that there is a class of
nonlinear collective and interacting spin models that can be simulated with current technology.

Nonlinear collective and interacting spin
models have long  endured as tractable, nonlinear quantum models with wide ranging relevance. Such models have appeared in
nuclear physics\cite{Ring1980}, laser physics\cite{Drummond}, condensed matter physics\cite{Mahan} and of course as a 
theoretical laboratory to investigate aspects of nonlinear field theories\cite{Kaku}. In many cases however the match between
model and experiment is only qualitative. In this paper I will show how some of these models may be directly simulated on a
linear ion trap with individual ion addressing as in the quantum computation architecture.

The interaction Hamiltonian for N ions interacting with the centre of mass vibrational mode can be controlled by using
different kinds of Raman laser pulses. A considerable variety of interactions has already been achieved
or proposed \cite{NIST,LANL,James}. Consider first the simplest interaction that does not change the vibrational mode
of the ions. Each ion is assumeds to be driven by a resonant laser field which couples two states, the ground state $|g\rangle$
and an excited state $|e\rangle$. The interaction Hamiltonian is 
\begin{equation}
H_I=-\frac{i}{2}\hbar\sum_{i=1}^N(\Omega_i\sigma^{(i)}_+-\Omega_i^*\sigma^{(i)}_-)
\end{equation}
where $\Omega_i$ is the effective Rabi frequency at the i'th ion and we have assumed the dipole and rotating wave
approximation as usual. The raising and lowering operators for each ion are defined by $\sigma_-=|g\rangle \langle e|$ and
$\sigma_+=|e\rangle\langle g|$.  If we now assume that each ion is driven by an
identical field and chose the phase appropriately, the interaction may be written as
\begin{equation}
H_I=\hbar\Omega \hat{J}_y
\label{rotation}
\end{equation}
where we have used  the definition of the collective spin operators,
\begin{equation}
\hat{J}_\alpha=\sum_{i=1}^N\sigma^{(i)}_\alpha
\end{equation}
where $\alpha=x,y,z$ and 
\begin{eqnarray}
\sigma^{(i)}_x & = & \frac{1}{2}(\sigma^{(i)}_++\sigma^{(i)}_-)\\
\sigma^{(i)}_y & = & -\frac{i}{2}(\sigma^{(i)}_+-\sigma^{(i)}_-)\\
\sigma^{(i)}_x & = & \frac{1}{2}(|e\rangle\langle e|-|g\rangle\langle g|)
\end{eqnarray}
The interaction Hamiltonian in Eq \ref{rotation} corresponds to a single collective spin of value $j=N/2$ precessing around
the $\hat{J}_y$ direction due to an applied field. By choosing the driving field on each ion to be the same we have imposed a
permutation symmetry in the ions reducing the dimension of the Hilbert space from $2^N$ to $2N+1$.  The eigenstates of
$\hat{J}_z$ may be taken as a basis in this reduced Hilbert space.  In ion trap quantum computers it is more usual to designate
the electronic states with a binary number as $|g\rangle=|0\rangle,\ \ |e\rangle=|1\rangle$. The product basis for  all N ions
is then specified by a single binary string, or the corresponding integer code if the ions can be ordered.  Each eigenstate,
$|j,m\rangle_z$,  of $\hat{J}_z$ is a degenerate eigenstate of the Hamming weight operator ( the sum of the number of ones in
a string)  on the binary strings labelling  the product basis  states in the $2^N$ dimensional Hilbert space of all possible
binary strings of length N. Collective spin models of this kind were considered many decades ago in quantum
optics\cite{Drummond} and are sometimes called Dicke models after the early work  on superradiance of Dicke\cite{Dicke}. In
much of that work however the collective spin underwent an irreversible decay. In the case of an ion trap model however we can
neglect such decays due to the long lifetimes of the excited states. However when the electronic and vibrational motion is
coupled heating of the vibrational centre-of-mass mode can induce irreversible dynamics in the collective spin variables.

The natural variable to measure is $\hat{J}_z$ as a direct determination of the state of each ion via shelving techniques will
give such a measurement. These measurements are highly efficient, approaching ideal projective measurements.   The
result of the measurement is a binary string which is an eigenstate of
$\hat{J}_z$.  Repeating such measurements it is possible to construct the distribution for $\hat{J}_z$ and corresponding
averages. Other components may also be measured by first using a collective rotation of the state of the ions.  

We now show how to realise nonlinear Hamiltonians using N trapped ions. By appropriate choice of Raman lasers it is possible
to realise the conditional displacement operator for the i'th ion\cite{Monroe1996,NIST} 
\begin{equation}
H=-i\hbar(\alpha_ia^\dagger-\alpha_i^* a)\sigma_z^{(i)}
\end{equation}
If the ion is in the excited (ground) state this Hamiltonian displaces the vibrational mode by a complex amplitude $\alpha$
($-\alpha$). In the case of N ions with each driven by identical Raman lasers, the total Hamiltonian is 
\begin{equation}
H=-i\hbar(\alpha a^\dagger-\alpha^* a)\hat{J}_z
\end{equation}
By an appropriate choice of Raman laser pulse  phases we can then implement the following sequence of unitary transformations
\begin{equation}
U_{NL}=e^{i\kappa_x\hat{X}\hat{J}_z}e^{i\kappa_p\hat{P}\hat{J}_z}e^{-i\kappa_x\hat{X}\hat{J}_z}e^{i\kappa_p\hat{P}\hat{J}_z}
\end{equation}
where $\hat{X}=(a+a^\dagger)/\sqrt{2},\ \hat{P}=-i(a-a^\dagger)/\sqrt{2}$. Noting that 
\begin{equation}
e^{i\kappa_p\hat{P}\hat{J}_z} \hat{X}e^{-i\kappa_p\hat{P}\hat{J}_z} =\hat{X}+\kappa_p\hat{J}_z
\end{equation}
it is easy to see that 
\begin{equation}
U_{NL}=e^{-i\theta\hat{J}_z^2}
\label{UNL}
\end{equation}
where $\theta=\kappa_x\kappa_p$ which is the unitary transformation generated by a nonlinear top Hamiltonian describing
precession around the $\hat{J}_z$ axis at a rate dependant on the $z$ component of angular momentum. Such nonlinear tops have
appeared in collective nuclear models\cite{Ring1980} and form the basis of a well known quantum chaotic system\cite{Haake91}.

It should be noted that the transformation in Eq(\ref{UNL}) contains no operators that act on the vibrational state. It is thus
completely independent of the vibrational state and it does not matter if the vibrational state is cooled to the ground state or
not. However Eq(\ref{UNL}) only holds if the heating of the vibrational mode can be neglected over the time it takes to
apply the conditional displacement operators. We discuss below what this implies for current experiments. 

In itself the unitary transformation in Eq (\ref{UNL}) can generate interesting states. For example if we begin with all the
ions in the ground state so that the collective spin state is initially $|j,-j\rangle_z$ and apply laser pulses to each
electronic transition according to the Hamiltonian in Eq (\ref{rotation}) for a time $T$ such that $\Omega T=\pi/2$ the
collective spin state is just the $\hat{J}_x$ eigenstate $|j,-j\rangle_x$. If we now apply the nonlinear unitary transformation
in Eq (\ref{UNL}) so that $\theta=\pi/2$ we find that the system evolves to the highly entangled state
\begin{equation}
|+\rangle=\frac{1}{\sqrt{2}}(e^{-i\pi/4}|j,-j\rangle_x+(-1)^je^{i\pi/4}|j,j\rangle_x)
\end{equation}
Such states have been considered by Bollinger et al.\cite{Bollinger} in the context of high precision frequency
measurements,and also by  Sanders\cite{Sanders}. They exhibit interference fringes for measurements of
$\hat{J}_z$. As noted above a measurement of $\hat{J}_z$ is easily made simply by reading out the state of each ion using
highly efficient fluorescence shelving techniques. This particular nonlinear model is a well known system for studying quantum
chaos, as we now discuss. 

The nonlinear top model was introduced by Haake\cite{Haake91,Sanders89} as a system that could exhibit chaos in the classical
limit on a compact phase space, and which could be  treated quantum mechanically with a finite Hilbert space. This removed the
necessity of truncating the Hilbert space and the possibility of thereby introducing spurious quantum features. The nonlinear
top is defined by the collective spin Hamiltonian,
\begin{equation} 
H = \frac{\kappa}{2 j \tau} \hat{J}_z^2 + p \hat{J}_y \sum_{n = -\infty}^\infty
\delta(t-n\tau),
\label{eq.top.hamiltonian}
\end{equation}
where $\tau$ is the duration between kicks, ${\bf \hat{J}} = (\hat{J}_x, \hat{J}_y, \hat{J}_z)$
is the angular momentum vector, and $\hat{J}^2=j(j+1)$ is a constant of the motion. 

As the Hamiltonian is time periodic the appropriate quantum description is via the Floquet operator
\begin{equation}
U = \exp\left(-i\frac{3}{2 j} \hat{J}_z^2\right)
\exp\left(-i\frac{\pi}{2} \hat{J}_y\right),   \label{eq.U.top}
\end{equation}
which takes a state from just before one kick to just before the next,
i.e.,
$\left|\psi\right> \longrightarrow U\left|\psi\right>$,
where $J_z$ and $J_y$ are the usual angular momentum operators, and
$j$ is the angular momentum quantum number. The first exponential,
$U_P = \exp\left(-i\frac{3}{2 j} \hat{J}_z^2\right )$,  describes the
precession about the $z$-axis, and the second, 
$U_K = \exp\left(-i\frac{\pi}{2} \hat{J}_y\right)$, describes the kick. 

The classical dynamics can be
reduced to a two dimensional map of points on a sphere of radius $j$ 
\cite{Sanders89}, and the angular momentum vector can be parameterised
in 
polar coordinates as
\begin{equation}
{\bf J} = j(\sin\Theta \cos\Phi, \sin\Theta \sin\Phi, \cos \Theta).
\end{equation}
The first term in the Hamiltonian (\ref{eq.top.hamiltonian}) 
describes a non-linear precession of 
the top about the z-axis, and the second term describes
periodic kicks around the y-axis.
The classical
map for $p = \pi /2$ and $\kappa  = 3$ has a mixed phase space with periodic elliptical fixed points and chaotic regions.

It is now clear that this model can be simulated by the
sequence of pulses in Eq ({\ref{UNL}) with appropriate values for the pulse area, together with a single linear rotation. This
presents the possibility of directly testing a number of ideas in the area of quantum chaos, particularly the idea of
hypersensitivity to perturbation introduced by Schack and Caves \cite{Caves94}. Of particular interest here is the
ability to very precisely simulate the measurement induced hypersensitivity discussed in \cite{Breslin99}. In that paper the
kicked top was subjected to a readout using a single spin that could be prepared in a variety of states. The interaction
between the top and the readout spin is described by
\begin{equation}
U_I = \exp\left(-i \mu J_y \sigma_z^{(R)}\right),
\end{equation}
where we regard one ion as set aside to do the readout and label it with a superscript. 
It is relatively straight forward to generate this interaction via the pulse sequence of conditional phase shifts
\begin{equation}
U_{NL}=e^{i\kappa_x\hat{X}\sigma_z^{(R)}}e^{i\kappa_p\hat{P}\sigma_z^{(R)}}e^{-i\kappa_x\hat{X}\hat{J}_z}e^{i\kappa_p\hat{P}\sigma_z^{(R)}}
\end{equation}
with $\mu=\kappa_x\kappa_p$. It is now possible to consider a long sequence of measurements made at the end of each nonlinear
kick and record the resulting binary strings of measurement results.  

Initial states of the kicked top can be easily be
prepared as coherent angular momentum states by appropriate linear rotations.   In the basis of orthonormal $\hat{J}_z\mbox{
eigenstates}$, and $\hat{{\bf J}}^2\left|j,m\right> = j(j+1)\left|j,m\right>$. 
the spin coherent states can be written as a rotation of the collective ground state\cite{Haake91,Arrechi72}
through the spherical polar angles $(\theta,\phi)$, 
\begin{equation}
|\gamma\rangle = \exp\left [i\theta(\hat{J}_x\sin\phi-\hat{J}_y\cos\phi))\right ]|j,-j\rangle
\label{eq.coherent}
\end{equation}
where $\gamma = e^{i\Phi} \tan\left(\frac{\Theta}{2}\right)$. This can be achieved by identical, appropriately phased pulses
on each ion separately. Initial states localised in either the regular or chaotic regions of the classical phase space may thus
be easily prepared. 

Using a sequence of conditional displacement operators that does distinguish different ions we can simulate
various interacting spin models. As interacting spins are required for general quantum logic gates, these models may be seen as
a way to perform quantum logical operations without first cooling the ions to the ground state of some collective vibrational
mode. 

Suppose for example we wish to simulate the interaction of two spins with the Hamiltonian
\begin{equation}
H_{int}=\hbar\chi\sigma_z^{(1)}\sigma_z^{(2)}
\end{equation}
The required pulse sequence is 
\begin{eqnarray}
U_{int} & = &
e^{i\kappa_x\hat{X}\sigma_z^{(1)}}e^{i\kappa_p\hat{P}\sigma_z^{(2)}}e^{-i\kappa_x\hat{X}\sigma_z^{(1)}}e^{i\kappa_p\hat{X}\sigma_z^{(2)}}\\
\nonumber & = & e^{-i\chi\sigma_z^{(1)}\sigma_z^{(2)}}
\end{eqnarray}

This transformation may be used together with single spin rotations to simulate a two spin transformation that is one of the
universal two qubit gates for quantum computation. For example the controlled phase shift operation
\begin{equation}
U_{cp}=e^{-i\pi|e\rangle_1\langle e|\otimes|e\rangle_2\langle e|}
\end{equation}
may be realised with $\chi=\pi$ as 
\begin{equation}
U_{cp}=e^{-i\frac{\pi}{2}\sigma_z^{(1)}}e^{-i\frac{\pi}{2}\sigma_z^{(2)}}U_{int}
\end{equation}
Once again this transformation does not depend on the vibrational state and so long as it is applied faster than the heating
rate of the collective vibrational mode it can describe the effective interaction between two qubits independent of the
vibrational mode. 

We have proposed a scheme, based on conditional displacements of a collective vibrational mode, to simulate  a variety of
nonlinear spin models using a linear ion trap in the quantum computing architecture and which does not require that the
collective vibrational mode be cooled to the ground state. However the scheme does require that the heating of the collective
vibrational mode is negligible over the time of the application of the Raman conditional displacement pulses. It does not
matter that the ion heats up between pulses. If the pulses were applied for times comparable to the heating times the pulse
sequences described above would not be defined by a product of unitary transformations but rather by the completely positive
maps which include the unitary part as well as the nonunitary heating part. Such maps provide a means to test various
thermodynamic limits of nonlinear spin models and will be discussed in a future paper. In current experiments the heating time
is estimated to be of the order of 1 ms, which is much shorter than the theoretically expected values that are as long as
seconds\cite{NIST}. The source of this heating is unclear but efforts are under way to eliminate it, so we can
expect heating times to eventually be sufficiently long to ignore. In current experiments however  the sequence of conditional
displacements would need to be applied on time scales of less than 1 ms. This is achievable using Raman pulses. We thus
conclude that simple collective and interacting spin models with a few spins are within reach of current ion trap quantum
computer experiments.   
 
\acknowledgements
I would like to thank Daniel James and David Wineland for useful discussions.

\end{document}